0

\documentclass[twocolumn,showpacs,preprintnumbers,pra,amsmath,amssymb]{revtex4}
\usepackage{graphicx}
\usepackage{bm}

\def\bea{\begin{eqnarray}}
\def\eea{\end{eqnarray}}

\begin{document}
\title{Optimal quantum state estimation with use of the no-signaling principle}
\author{Yeong-Deok Han $^{1}$, Joonwoo Bae $^{2}$, Xiang-Bin Wang $^{3}$, and Won-Young Hwang $^{4}$
\footnote{wyhwang@jnu.ac.kr}}
\affiliation{$^{1}$ Department of Game Contents, Woosuk University, Wanju, Cheonbuk 565-701, Republic of Korea \\$^{2}$School of Computational Sciences, Korea Institute for Advanced Study, Seoul 130-012, Republic of Korea \\ $^{3}$
Department of Physics and the Key Laboratory of Atomic and Nanosciences, Ministry of Education, Tsinghua University, Beijing 100084, China.
\\ $^{4}$Department of Physics Education, Chonnam National University, Gwangju 500-757, Republic of Korea}
\begin{abstract}
A simple derivation of the optimal state estimation of a quantum bit was obtained by using the no-signaling principle. In particular, the no-signaling principle determines a unique form of the guessing probability independently of figures of merit, such as the fidelity or information gain. This proves that the optimal estimation for a quantum bit can be achieved by the same measurement for almost all figures of merit.

\end{abstract}
\maketitle
\section{Introduction}\label{sec:intro}
Special relativity contains the no-signaling principle whereby no information can be transferred faster than light. Quantum nonlocality appears to contradict the no-signaling principle. However, quantum nonlocality and the no-signaling principle are in "peaceful coexistence" \cite{Shi83}.

For such coexistence, the no-signaling principle places constraints on the behavior of quantum systems. Interestingly, the bounds obtained by the no-signaling constraint are the same as those obtained using purely quantum mechanical methods. For example, there is optimal quantum cloning \cite{Gis98}, optimal unambiguous state discrimination \cite{Bar02}, minimal error state discrimination \cite{Hwa05,Bae08,Hwa10}, and maximum confidence state discrimination \cite{Cro08}.

The purpose of this article is to add one to the list. The topic of this study is the optimal state estimation for a single quantum bit (qubit). Massar and Popescu \cite{Mas95} reported that the maximum average fidelity for a single qubit estimation is $2/3$. Using only spatial symmetry, on the other hand, Han \cite{Han05} suggested a way to derive optimal quantum direction transfer \cite{Bag00}, which is the same as the state estimation problem, for the case of a single qubit.
This article shows how the optimal estimation of the single qubit can be obtained simply by using the no-signaling principle. Moreover, the results show that, for any figure of merit, the guessing-probability (distributions) has the same form, $ A \cos^2 (\theta/2)+ B \sin^2 (\theta/2) $. Here $A,B$ are constants and $\theta$ is the angle between the Bloch vector of the prepared qubit and that of the guessed one. This result actually confirms the suggestion in Refs. \cite{Bag00,Tar99} that the optimal measurements are the same for any figure of merit.
\section{quantum state estimation}
First the procedures of a quantum state estimation are described more precisely. A player, Alice, randomly chooses a direction in three-dimensional space. That is, she chooses a unit vector $\hat{r}$ with an isotropic probability distribution. Then she prepares a qubit in the pure state with its Bloch vector $\hat{r}$; namely,
\begin{equation}
\rho(\hat{r}) = \frac{1}{2} (\openone +\hat{r} \cdot \vec{\sigma})= |\hat{r}\rangle \langle \hat{r}|.
\label{A}
\end{equation}
Here, $\hat{r} =(\sin\theta\cos\varphi,\sin\theta\sin\varphi,\cos\theta)$ and $\vec{\sigma}=(\sigma_x, \sigma_y, \sigma_z)$, where $\sigma_x, \sigma_y, \sigma_z$ are Pauli operators. Alice then sends a qubit $\rho(\hat{r})$ to another player, Bob. He knows all of Alice's procedures but does not know the identity of the qubit. He then guesses the identity of the qubit using all possible means including quantum measurement of the qubit. Bob's figure of merit (or score) is a function of the state of the sent qubit and that of the guessed qubit. Normally, the closer the two states are, the higher the figure of merit is. The commonly used figures of merit are fidelity and information gain  \cite{Tar99,Bag00}.
Bob's task is to obtain the maximal figures of merit, on average.

Bob's maximal average fidelity was reported to be  $2/3$ \cite{Mas95}. Bob's strategy for achieving the maximum is simple \cite{Mas95}: He  randomly chooses a unit vector $\hat{n}$ and performs a measurement $S_{\hat{n}}$ whose bases are $\rho(\hat{n})= |\hat{n}\rangle \langle \hat{n}|$ and $\rho(-\hat{n})= |-\hat{n}\rangle \langle -\hat{n}|$ on the input qubit. Physically, $S_{\hat{n}}$ corresponds to the Stern-Gerlach measurement in the $\hat{n}$ direction if the qubit is in the spin of a particle. Next, if the measurement outcome is $\rho(\hat{n})$ [$\rho(-\hat{n})$], he estimates that Alice has sent a qubit in the $\rho(\hat{n})$ [$\rho(-\hat{n})$] state. Consider the guessing-probability $P(\hat{m}|\hat{r})$. Here, $P(\hat{m}|\hat{r}) \hspace{1mm} d\Omega$ is the probability that an outcome $\rho(\hat{r}^{\prime})$ with a unit vector $\hat{r}^{\prime}$ around $\hat{m}$ within a solid angle $d\Omega$ is obtained for an input qubit $\rho(\hat{r})$. It is not difficult to see that
\begin{equation}
P(\hat{m}|\hat{r})=\frac{1}{2\pi}\cos^2\frac{\theta}{2} \hspace{2mm} \propto |\langle\hat{m}|\hat{r}\rangle|^2,
\label{A-2}
\end{equation}
where $\theta$ is the angle between  $\hat{m}$ and $\hat{r}$.
\section{guessing-probability has a unique form.}
This section introduces a communication scenario between two remotely separated participants, Alice and Bob. To the communication scenario, quantum state estimation can be incorporated, as in the case of the minimal error state discrimination in Refs.\cite{Hwa05,Bae08,Hwa10}. Suppose Alice and Bob share many copies of an entangled state:
\begin{equation}
|\psi\rangle= \sqrt{p} |0\rangle_A |\hat{z}\rangle_B +\sqrt{1-p} |1\rangle_A |-\hat{z}\rangle_B,
\label{B}
\end{equation}
 where $|0\rangle$ and $|1\rangle$ are two orthogonal states of a qubit, and $A$ and $B$ denote Alice and Bob, respectively. If Alice performs a measurement in the \{$|0\rangle$, $|1\rangle$\} basis, Bob is given a mixture of $|\hat{z}\rangle \langle\hat{z}|$ and $|-\hat{z}\rangle \langle -\hat{z}|$ with the respective probabilities $p$ and $1-p$. Bob's density operator is then given by,
\begin{equation}
\label{C}
\rho_B= p |\hat{z}\rangle \langle\hat{z}|+ (1-p) |-\hat{z}\rangle \langle -\hat{z}|= \frac{1}{2} \{\openone +\vec{r}_B\cdot \vec{\sigma}\},
\end{equation}
where $\vec{r}_B= p\hat{z}+(1-p)(-\hat{z})$ (see Fig.1).
\begin{figure}
  \includegraphics[width=7cm]{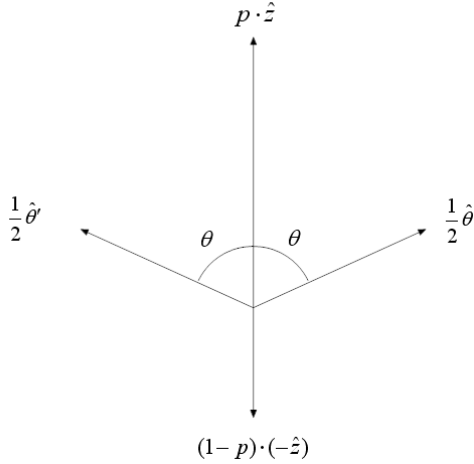}
 \caption{ Note that $p\hat{z}+(1-p)(-\hat{z})=(1/2)\hat{\theta}+(1/2)(\hat{\theta}^{\prime})$}\label{Fig-1}
\end{figure}
Note that the Bloch vector of a mixture is given by the sum of Bloch vectors of the pure states constituting the mixture with the corresponding probabilities as weighting factors. Consider Bob's Bloch vector, $\vec{r}_B =p\hat{z}+(1-p)(-\hat{z})$. Then consider a different decomposition of  $\vec{r}_B$;  $\vec{r}_B= (1/2)\hat{\theta}+(1/2)(\hat{\theta}^{\prime})$. Here $\hat{\theta}=(0,\sin\theta,\cos\theta), \hat{\theta}^{\prime}=(0,-\sin\theta,\cos\theta)$, and
\begin{equation}
\label{C-2}
\cos\theta=p-(1-p)=2p-1.
\end{equation}
This means that
\begin{equation}
\label{D}
\rho_B= \frac{1}{2} |\hat{\theta}\rangle \langle\hat{\theta}|+  \frac{1}{2} |\hat{\theta}^{\prime}\rangle \langle \hat{\theta}^{\prime}|= \frac{1}{2} \{\openone +\vec{r}_B\cdot \vec{\sigma}\}.
\end{equation}
However, according to the Gisin-Hughston-Jozsa-Wootters theorem \cite{Gis89,Hug93},
Alice can generate any decomposition of Bob's mixture by measuring her qubit on an appropriate basis. Therefore, in this case, either decomposition of Eq. (\ref{C}) or that of Eq. (\ref{D}) can be generated by Alice. This suggests that the entangled state can be written as follows:
\begin{equation}
|\psi\rangle= \frac{1}{\sqrt{2}} |0^{\prime}\rangle_A |\hat{\theta}\rangle_B +\frac{1}{\sqrt{2}} |1^{\prime}\rangle_A |\hat{\theta}^{\prime}\rangle_B,
\label{E}
\end{equation}
where $\{|0^{\prime}\rangle, |1^{\prime}\rangle\}$ is another orthogonal basis.
Therefore, Alice can generate a decomposition of Eq. (\ref{C}) (that of Eq. (\ref{D})) by measuring her qubit on a $\{|0\rangle, |1\rangle\}$ ($\{|0^{\prime}\rangle, |1^{\prime}\rangle\}$) basis.

However, if Bob can discriminate between the two decompositions, they can perform  faster than light communication. If Alice wants to send a message $0$ ($1$), she repeatedly performs the measurement on the $\{|0\rangle, |1\rangle\}$ ($\{|0^{\prime}\rangle, |1^{\prime}\rangle\}$) basis. The decomposition of Eq. (\ref{C}) [Eq. (\ref{D})] is then generated at Bob's site. Bob can read the message by discriminating the two decompositions.

Now let us show that the guessing probability has a unique form using the no-signaling principle. As described above, in the quantum state estimation, Bob's task is to estimate the input with maximal figure of merit, on average. He is allowed to use all possible means including classical and quantum computers, and even humans. Consider a "black-box", a "(quantum) state estimator," which includes everything needed for the estimation inside. For an input $\rho(\hat{r})$, the quantum-state estimator just gives an outcome, which is its optimal guess, $\hat{m}$.
Faster-than-light communication is possible unless the
guessing-probability $P(\hat{m}|\hat{r})$ is in a form
$A\sin^2(\theta/2)+ B\cos^2(\theta/2)$, where $\theta$ is angle between $\hat{z}$ and $\hat{\theta}$.
Because the two decompositions cannot be discriminated, it must be
\begin{equation}
\frac{1}{2} P(\hat{m}|\hat{\theta})+ \frac{1}{2}P(\hat{m}|\hat{\theta}^{\prime})=
p P(\hat{m}|\hat{z})+(1-p) P(\hat{m}|-\hat{z}).
\label{F2}
\end{equation}
for all directions $\hat{m}$.

At this stage, it is assumed that the estimator is isotropic; that is,
the guessing probability $P(\hat{m}|\hat{r})$ depends
dependent only on the angle between $\hat{m}$ and $\hat{r}$.

To obtain the functional form of $P(\hat{m}|\hat{\theta})$ most simply,
we consider the $\hat{m}=\hat{z}$ case
\begin{equation}
\frac{1}{2} P(\hat{z}|\hat{\theta})+ \frac{1}{2}P(\hat{z}|\hat{\theta}^{\prime})=
p P(\hat{z}|\hat{z})+(1-p) P(\hat{z}|-\hat{z}).
\label{F}
\end{equation}

If Eq. (\ref{F}) is not satisfied, Bob can discriminate the two decompositions by only observing how frequently the state estimator gives the outcome $\hat{z}$. More precisely, Bob counts the frequency that the state estimator gives the outcomes
 $\rho(\hat{r}^{\prime})$ with unit vector $\hat{r}^{\prime}$ around $\hat{z}$ within the  solid angle $d\Omega$.
In the case of the decomposition of Eq. (\ref{C}), we can see that the frequency is $\{p P(\hat{z}|\hat{z})+(1-p) P(\hat{z}|-\hat{z})\} d\Omega$. In the case of the decomposition of Eq.(\ref{D}), the frequency is $\{(1/2) P(\hat{z}|\hat{\theta})+ (1/2)P(\hat{z}|\hat{\theta}^{\prime})\} d\Omega$.

However, using the isotropy assumed above, we have
\begin{equation}
P(\hat{z}|\hat{\theta})=P(\hat{z}|\hat{\theta}^{\prime}).
\label{G}
\end{equation}
By Eqs. (\ref{C-2}), (\ref{F}) and (\ref{G}), and setting $P(\hat{z}|\hat{z})\equiv A$ and $P(\hat{z}|-\hat{z})\equiv B$,  we obtain
\begin{equation}
P(\hat{z}|\hat{\theta})= A \cos^2 \frac{\theta}{2}+ B \sin^2 \frac{\theta}{2},
\label{H}
\end{equation}
where $\theta$ is the angle between $\hat{z}$ and $\hat{\theta}$.

We can also rewrite it as follows:
\begin{equation}
P(\hat{z}|\hat{\theta})= \alpha + \beta \cos \theta = \alpha + \beta \hat{z} \cdot \hat{\theta},
\label{H2}
\end{equation}
where $\alpha= (A+B)/2$ and $\beta= (A-B)/2$.
 Using the isotropy assumed above, this result can be generalized for any direction $\hat{m}$ and $\hat{\theta}$:
\begin{equation}
P(\hat{m}|\hat{\theta})= \alpha + \beta \hat{m} \cdot \hat{\theta}.
\label{H3}
\end{equation}

Now, it is easy to show that this functional form generally satisfies Eq.(\ref{F2}).
Therefore, the general form of the guessing probability for state $\rho(\hat{\theta})$ can be expressed as
\begin{equation}
P(\hat{m}|\hat{\theta})= A \cos^2 \frac{\theta}{2}+ B \sin^2 \frac{\theta}{2},
\label{H}
\end{equation}
where $\theta$ is the angle between $\hat{m}$ and $\hat{\theta}$.


Interestingly, the guessing-probability has a unique form regardless of the figure of merit. If there is only a single guessing-probability, there is nothing to optimize. However, there are still infinitely many guessing-probabilities depending on the constants, $A$ and $B$. Therefore, it should be optimized. When the figure of merit is fidelity, it is easy to see that it is optimized when $B=0$, obtaining $P(\hat{m}|\hat{\theta})= A \cos^2 (\theta/2)$. The actual measurement strategy that achieves the optimal one is the simple strategy described in Sec II.
We recover Eq. (\ref{A-2}) after normalization. When the figure of merit is the information gain, it is optimized when either $B=0$ or $A=0$. The guessing probability in the former case is the same as the one when the figure of merit is fidelity. Therefore, the optimal measurements are the same in this case. However, the guessing probability in the latter case is reversed. However, the reversed case can be realized by the state estimator used in the former case. That is, $-\hat{r}$ is adopted as the true outcome when an outcome $\hat{r}$ is given. Hence, in the latter case, the optimal measurement is the same.
It can be expected that the optimal guessing probability is the same for other figure of merits. Almost all figures of merit have a property that the figure of merit increases with decreasing $\theta$. Provided the property is satisfied, the optimal guessing probability can be obtained when $B=0$. If the guessing probability is the same, the optimal measurement would also be the same.
\section{conclusion}
A simple derivation of the optimal quantum state estimation of a qubit was obtained  using the no-signaling principle. In particular, the no-signaling principle determines a unique form of the guessing probability, independent of figures of merit, such as the fidelity or information gain. An optimal guessing probability with the unique form can be realized using a simple actual measurement strategy. This proves that the optimal estimation for a qubit can be achieved by the same measurement for all figures of merit, provided the figure of merit satisfies the property that the figure of merit increases with decreasing $\theta$.
\section*{Acknowledgement}
This study was financially supported by Woosuk University.
This study was supported by Basic Science Research Program through the National Research Foundation of Korea (NRF) funded by the Ministry of Education, Science and Technology (2010-0007208 and KRF-2008-313-C00185).
This study was financially supported  by Chonnam National University, 2009.

\end{document}